# A Hybrid Ensemble Learning Framework for Image-Based Solar Panel Classification

Vivek Tetarwal
*Central Research Laboratory*
*Bharat Electronics Limited*
Ghaziabad, India
tetarwalvivek1991@gmail.com

Sandeep Kumar
*Central Research Laboratory*
*Bharat Electronics Limited*
Ghaziabad, India
sann.kaushik@gmail.com

*Abstract*— The installation of solar energy systems is on the rise, and therefore, appropriate maintenance techniques are required to be used in order to maintain maximum performance levels. One of the major challenges is the automated discrimination between clean and dirty solar panels. This paper presents a novel Dual Ensemble Neural Network (DENN) to classify solar panels using image-based features. The suggested approach utilizes the advantages offered by various ensemble models by integrating them into a dual framework, aimed at improving both classification accuracy and robustness. The DENN model is evaluated in comparison to current ensemble methods, showcasing its superior performance across a range of assessment metrics. The proposed approach performs the best compared to other methods and reaches state-of-the-art accuracy on experimental results for the Deep Solar Eye dataset, effectively serving predictive maintenance purposes in solar energy systems. It reveals the potential of hybrid ensemble learning techniques to further advance the prospects of automated solar panel inspections as a scalable solution to real-world challenges.

## I. INTRODUCTION

Solar energy is a basic component of sustainable energy frameworks, providing a clean and renewable source to meet the ever-growing global energy demands. However, the efficiency of solar panels in operation must be ensured to maximize their energy generation. One of the most significant factors affecting efficiency is soiling, which results from the accumulation of dust, dirt, and other particulates on panel surfaces. Accurate and automated detection of soiled solar panels is essential for timely cleaning and maintenance, helping to minimize energy losses and reduce operational costs.

Traditionally, methods for assessing the cleanliness of solar panels relied on manual inspection or sensor-based techniques, which are resource-intensive and impractical for large-scale installations. Recent advances in computer vision and machine learning have enabled image-based analysis, providing a non-invasive, cost-effective alternative. Deep learning models, particularly convolutional neural networks (CNNs), have demonstrated remarkable performance in classifying solar panels based on their soiling status [1], [2].

Several deep learning-based approaches have been proposed to tackle efficiency issues caused by dust accumulation. For example, models utilizing Inceptionv3, VGG16, and custom CNN architectures have been developed for soiling detection [3]. SolNet, a novel CNN-based model, was specifically designed to detect dust accumulation on solar panels [4]. AI-driven predictive maintenance strategies employing deep learning have been introduced to monitor soiling while reducing computational costs [5]. Additionally, machine learning models using visible spectrum data have been evaluated, with CNNs achieving the highest test performance [6]. Other approaches have leveraged RGB images and environmental data to predict PV module power loss due to soiling and shading [7], while deep learning-based probabilistic methods have been used to estimate power loss from surveillance images [8]. Real-time detection frameworks, such as YOLOv5, have also been applied to PV panel soiling classification [9]. However, these architectures often face challenges in dealing with the multiplicity of environmental factors, including lighting changes and occlusions, which may affect their performance.

Ensemble learning has emerged as a promising approach to improve classification accuracy by leveraging the strengths of multiple models. Techniques such as bagging and boosting, exemplified by Random Forest and Gradient Boosting Machines, have outperformed single models in solar panel classification tasks [10], [11]. More recently, ensembles of neural networks with varying architectures and feature representations have been explored to enhance accuracy in complex image classification scenarios [12], [13]. Despite these advancements, existing ensemble methods often struggle to maintain a balance between robustness and computational efficiency.

In this work, we have proposed a novel ensemble-based approach namely Dual Ensemble Neural Network (DENN) for classifying solar panels as clean or soiled based on image-derived features. Our approach combines two complementary techniques within an architecture, effectively combining the strengths of diverse strategies. This dual-ensemble design enhances generalization across varied datasets and environmental conditions while mitigating the limitations of conventional single-ensemble techniques.

To validate the proposed approach, we conducted extensive experiments using the Deep Solar Eye dataset [14], a publicly available dataset which contains RGB images of solar panels under different soiling conditions as illustrated in Fig. 1. The dataset is particularly useful for tasks related to soiling detection and power loss prediction in photovoltaic modules. Experimental results show that DENN outperforms traditional ensemble techniques significantly, providing higher accuracy and soiling detection robustness.

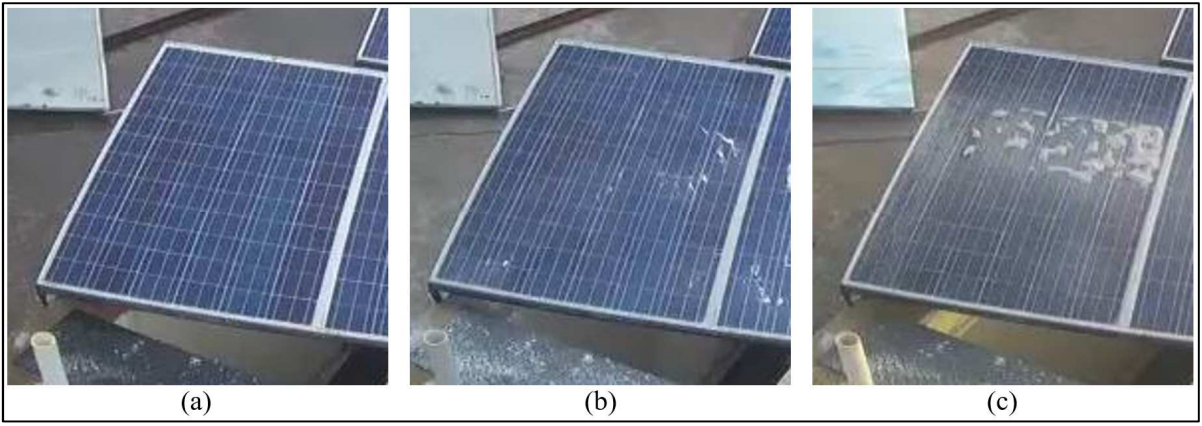

Fig. 1. Images from DeepSolarEye Dataset [14]: (a) Clean panel; (b) & (c) Soiled panels

The organization of the paper is as follows: Section II describes the detailed methodology, including feature extraction and ensemble techniques. In Section III, the experimental results are presented and analysed, highlighting the comparative performance of various methods. Finally, Section IV concludes the paper and outlines potential future directions for improving solar panel defect detection.

## II. METHODOLOGY

The objective of this research is to develop and evaluate the DENN framework for the detection of soiled solar panels using deep feature extraction and ensemble learning techniques. The methodology combines state-of-the-art computer vision methods to extract meaningful representations from solar panel images and employs heterogeneous ensemble methods to improve classification performance. The process involves feature extraction using pretrained convolutional neural networks (CNNs), followed by training and testing various ensemble-based models. A detailed overview of the methodology is provided in the following subsections.

### A. Dataset

The data set used in this research work consists of 45,721 solar panel images, each with unique power loss values. The images are classified into two categories on the basis of the power loss values: soiled and clean. To use the data set for feature extraction and then classification, the data set is divided into two sets: training, 80%, and testing, 20%. This split ensures that the models are trained on a substantial portion of the data while maintaining distinct sets for hyperparameter tuning and performance evaluation.

### B. Feature Extraction

Pretrained ResNet-50, is employed to extract hierarchical features from the solar panel images. The fully connected layers of ResNet-50 are removed to retain only the convolutional base. The solar panel images are passed through the modified ResNet-50, and the output feature maps are flattened into one-dimensional vectors. The extracted feature vectors are then paired with their corresponding labels for classification tasks.

### C. Preprocessing

In order to mitigate class imbalance, present within the dataset, the Synthetic Minority Oversampling Technique (SMOTE) was utilized. It was applied to create synthetic samples for the minority class through interpolation between existing samples. In this way, the dataset would be balanced without creating overfitting, since there was no simple duplication of existing samples.

### D. System Architecture

The dual ensemble architecture of the proposed method is as illustrated in Fig. 2.

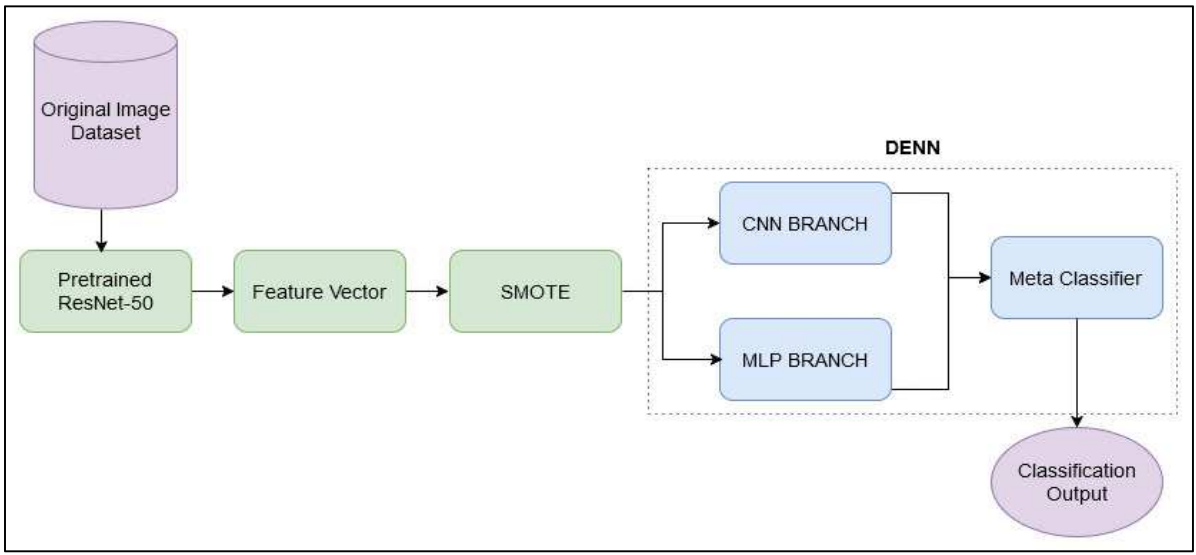

Fig. 2. System Block Diagram

The Dual Ensemble Neural Network (DENN) is an advanced ensemble technique that integrates two independent feature extraction branches—a CNN-based branch and an MLP-based branch—to enhance classification performance by leveraging complementary feature representations. Mathematically, given an input feature vector $X \in R^d$, DENN splits it into two equal parts: $X_{CNN} \in R^{d/2}$ for the CNN branch and $X_{MLP} \in R^{d/2}$ for the MLP branch. Each branch applies a linear transformation followed by a ReLU activation function, yielding feature representations $F_{CNN}$ and $F_{MLP}$, computed as:

$$F_{CNN} = RELU(W_{CNN}X_{CNN} + b_{CNN}) \quad (1)$$

$$F_{MLP} = RELU(W_{MLP}X_{MLP} + b_{MLP}) \quad (2)$$

where $W_{CNN}, b_{CNN}, W_{MLP}, b_{MLP}$ are the trainable weights and biases of the respective branches. The extracted feature vectors are then concatenated to form a unified feature representation:

$$F_{dual} = [F_{CNN}; F_{MLP}] \in R^{256} \quad (3)$$

This fused feature vector is then passed through a meta-classifier, which applies another linear transformation to predict the final class probabilities:

$$\hat{Y} = \sigma(W_{meta}F_{dual} + b_{meta}) \quad (4)$$

where $W_{meta}, b_{meta}$ represent the meta-classifier's parameters, and $\sigma$ denotes the activation function (e.g., SoftMax for multi-class classification). The model is trained using cross-entropy loss:

$$L = - \sum_{i=1}^{N} Y_i \log(\hat{Y}_i) \quad (5)$$

where $Y_i$ is the ground truth label. By combining CNN and MLP feature extraction, DENN enhances model robustness, effectively leveraging both spatial and abstract representations, leading to improved classification accuracy and generalization.

### III. EXPERIMENTAL SETUP

The model was implemented with PyTorch and optimized by the Adam optimizer, which has efficient and adaptive learning during training. Training was carried out for 30 epochs with periodic testing to monitor how well the model was performing. Finally, it was tested against a held-out test set in order to ensure generalization of the model.

During the inference phase, the SoftMax layer was used to compute the probabilities related to each class, thus giving a quantifiable confidence level for every possible label. The label that had the highest probability was selected as the final prediction. This methodology not only ensures accurate classification but also provides interpretability through probabilistic results. In short, DENN successfully combines the strengths of CNNs and MLPs, thus leading to a robust ensemble framework for the classification of solar panels. To establish the performance of DENN, other ensemble methods were applied exclusively for comparative purposes. All these ensemble methods depend on one or another strategy for combining the best of different models to produce sound and accurate predictions.

A basic ensemble method applied in this case is bagging, especially for Random Forest. It consists of training a large number of decision trees on random subsets of data and combining their predictions using majority vote. The code was run using the RandomForestClassifier from the sklearn library, which can avoid overfitting and hence improves generalization. Boosting (XGBoost) is sequentially derived by iteratively training models to correct the mistakes of their immediate predecessors. XGBoost, derived using the XGBClassifier, effectively carries out this process with gradient boosting algorithms with greater accuracy and computational speed.

The Voting Ensemble combines predictions from three different classifiers, namely Logistic Regression, Random Forest, and Support Vector Machines (SVM). Soft voting is used here; probabilistic predictions from each model are aggregated to make the final decision. Thus, the module VotingClassifier from sklearn was used for this purpose. Cascading represents a stratified ensemble approach whereby the predictions produced by a level-1 model, namely Random Forest, are used as supplementary features for a level-2 meta-model, namely Logistic Regression. The overall idea is that the method can increase accuracy by combining the original features with intermediate predictions.

Another technique of layered ensemble is called blending, which further splits the training data into two subsets. The base models are Random Forest and Logistic Regression, which are trained on the first subset, and the meta-model Logistic Regression trains on the predictions made by the former based on the second subset. It is a two-stage process that captures the strength of the model. The Dual Bagging & Boosting technique combines the capabilities of Random Forest (bagging) and XGBoost (boosting) by averaging their predicted probabilities. The final predictions are derived from the combined probabilities, ensuring a balance between the diversity of bagging and the sequential refinement of boosting.

Overall, Dynamic Ensemble encompasses the practice of synthesizing predictions coming from different models based on dynamic averages of estimated probabilities. Finally, the output is a conclusive classification based on some threshold that's applied on averaged probabilities. Since the DENN architecture caters to various models with differences in strengths, its adaptability and resiliency enhance.

Evaluation approaches for ensemble-based techniques reveal a series of possible comparisons between different configurations and effectiveness to benefit the DENN architecture under review.

### IV. RESULTS AND DISCUSSION

In this section, we have analyzed the outcomes of the experimentation conducted in this study. Python and the PyTorch framework on a system equipped with an NVIDIA RTX A5000 GPU are utilized. The results are evaluated in terms of various performance metrices such as precision, recall, accuracy, F1-Score and G-Mean.

The proposed DENN model showed excellent training performance, which was reflected by the steady loss reduction over 30 epochs. The loss dropped from 0.6100 in the second epoch to 0.3173 in the final epoch, signifying effective learning and optimization. The performance of the model was further validated by using confusion matrices and key evaluation metrics. As illustrated in Fig. 3, the confusion matrix for the DENN clearly shows high true positive and true negative rates, while the misclassification is minimal, which supports the capability of the model to handle complex patterns and classify samples precisely.

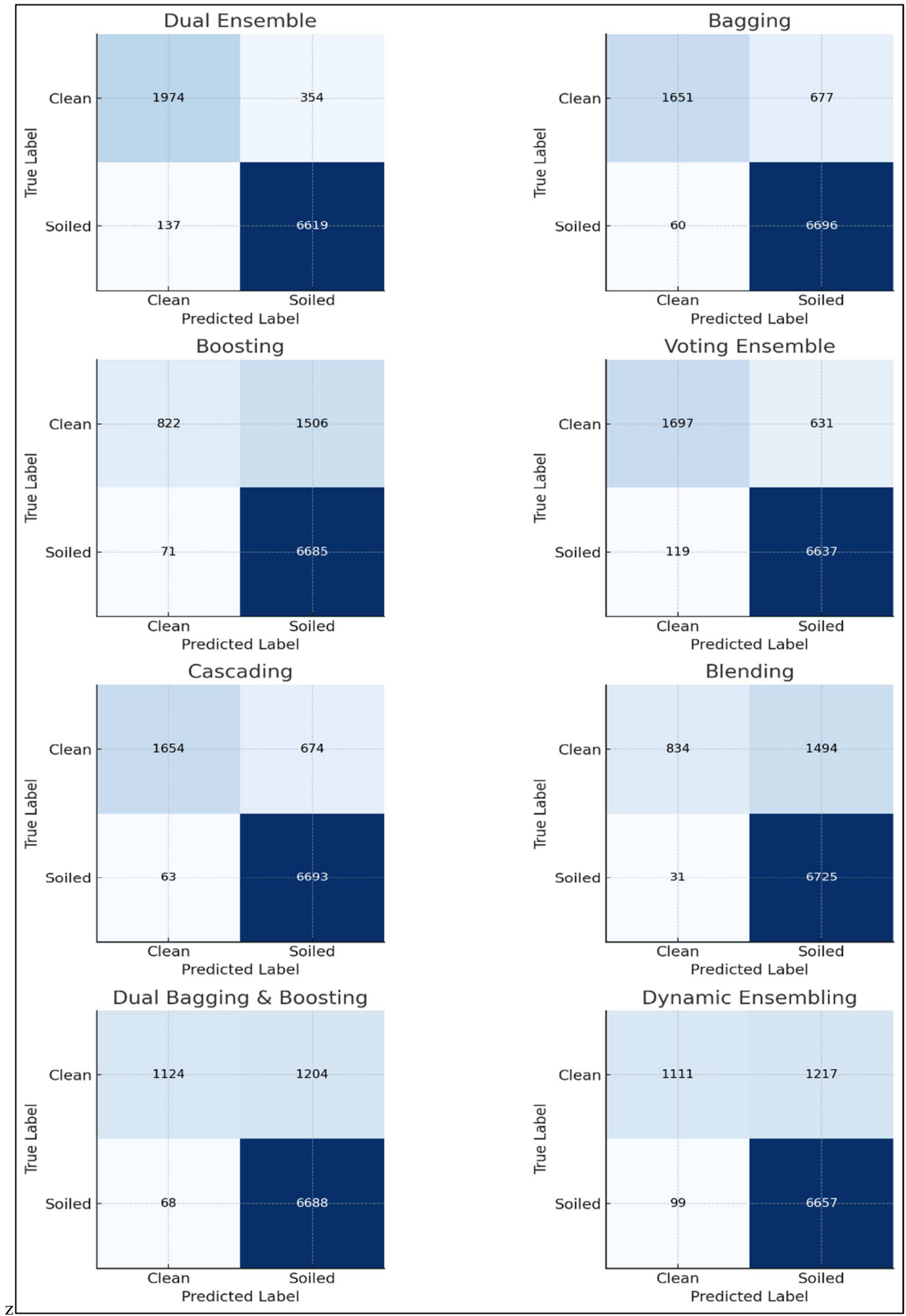

Fig. 3. Confusion Matrices for DENN and other ensemble methods

The overall comparison was given by comparing the DENN with other ensemble techniques such as Bagging, Boosting, Voting Ensemble, Cascading, Blending, Dual Bagging & Boosting, and Dynamic Ensemble.

Table 1 summarizes the performance metrics for all methods, including accuracy, precision, recall, F1-score, and geometric mean (G-Mean). The DENN outperformed these methods, achieving the highest accuracy (94.59%), precision (94.56%), recall (94.59%), and F1-score (94.51%). The G-Mean value of 0.8956 further emphasized the DENN's robustness, particularly in handling imbalanced datasets.

TABLE I. PERFORMANCE METRICS OF DENN AND OTHER ENSEMBLE TECHNIQUES

| Method | Accuracy | Precision | Recall | F1-Score | G-Mean |
|---|---|---|---|---|---|
| **Bagging [15]** | 0.918868 | 0.922723 | 0.918868 | 0.914443 | 0.807251 |
| **Boosting [16]** | 0.826398 | 0.842883 | 0.826398 | 0.79606 | 0.540181 |
| **Voting [17]** | 0.917437 | 0.918637 | 0.917437 | 0.913844 | 0.817782 |
| **Cascading [18]** | 0.918868 | 0.922554 | 0.918868 | 0.914496 | 0.807984 |
| **Blending [19]** | 0.832122 | 0.855626 | 0.832122 | 0.801863 | 0.545991 |
| **Dual Bagging & Boosting [20]** | 0.859974 | 0.871918 | 0.859974 | 0.842808 | 0.644368 |
| **Dynamic Ensembling [21]** | 0.855130 | 0.864082 | 0.85513 | 0.837776 | 0.638825 |
| **DENN** | **0.945949** | **0.945611** | **0.945949** | **0.945055** | **0.895604** |

From the baseline, it can be observed that both Bagging and Cascading performed close to each other at 91.89%. On the other hand, the performance was lower for both Boosting and Blending; with their accuracies at 82.64% and 83.21% while their respective F1 scores were below 80.2%. Dynamic Ensemble and Dual Bagging & Boosting achieved the mid-range with an accuracy score of 85.51% and 85.99%.

Fig. 4 shows the comparison of different ensemble learning methods based on four metrics namely Accuracy, Precision, F1-Score, and G-Mean by means of a radar chart. The legend indicates various ensemble techniques, including Dual Ensemble, Bagging, Boosting, Voting Ensemble, Cascading, Blending, Dual Bagging & Boosting, and Dynamic Ensembling.

The chart highlights the performance differences among these methods, with Dual Ensemble showing strong results across all metrics.

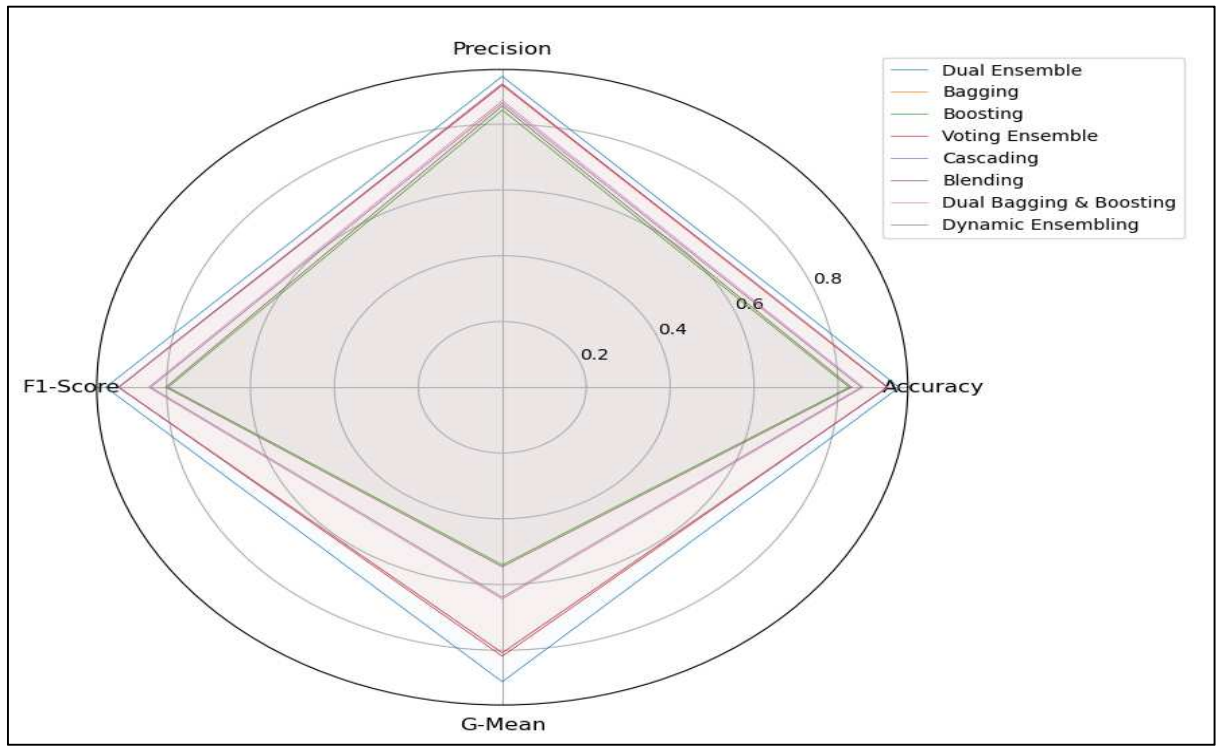

Fig. 4. Radar Chart showing different metrics for different ensemble methods

The results confirm that the DENN architecture, which integrates CNN and MLP branches with a meta-classifier, leverages both feature patterns and non-linear interactions to deliver state-of-the-art performance. Such findings set a benchmark for solar panel classification tasks, demonstrating the potential of advanced ensemble methods in similar domains. The DENN is superior to traditional ensemble methods because it performs feature-level ensemble rather than decision-level ensemble, leading to a more robust and generalizable model. Unlike methods such as Bagging, Boosting, Voting, and Stacking, which aggregate multiple models at the final classification stage, DENN integrates two distinct feature extraction branches: a CNN-based branch and an MLP-based branch, before making predictions. The CNN branch captures spatial and hierarchical features, which are crucial for structured data like images, while the MLP branch extracts abstract and high-dimensional representations, making it effective for general feature processing. Mathematically, given an input feature vector XXX, DENN splits it into two parts, processes them through separate neural networks, and concatenates the outputs:

$$F_{dual} = [F_{CNN}; F_{MLP}] \in R^{256} \quad (6)$$

This complementary feature representation enables the classifier to leverage richer information than models trained on a single feature extraction method. Traditional ensemble methods like Bagging and Boosting operate by training multiple models on variations of the dataset but do not enhance the underlying feature representation. Even Stacking Ensembles, which use a meta-classifier to combine multiple base models, fail to improve feature extraction at an earlier stage. DENN, however, constructs a more informative feature space, leading to improved learning and better classification accuracy. By performing ensemble at the feature level, DENN inherently reduces both bias and variance, making it theoretically more stable, data-efficient, and generalizable compared to decision-level ensemble methods.

## V. Conclusions and Future Work

This study proposed a classification method DENN for distinguishing clean solar panels from soiled ones, based on dual ensemble NN methods. DENN proves to be the best performing model with an accuracy of 94.59%, precision of 94.56%, recall of 94.59%, and F1-score of 94.50% that has an optimum G-Mean value of 0.89. Traditional ensemble techniques such as Bagging and Cascading performed competitively and reached higher than 91% accuracy, whereas methods such as Boosting and Blending, although of better performance, exhibit their sensitivity toward noise or some samples being miss-classified. Such detail in the matrices and comparisons emphasizes the need for combining ensemble approaches to achieve powerful results. While applying SMOTE during preprocessing eliminated the effects of class imbalance effectively, it contributed to the high performance of each classifier.

Although the proposed models obtain good classification performance, there are many directions for further improvement and exploration. Future studies could explore the use of transformer-based models to improve feature extraction from solar panel images. Adding metadata such as weather conditions, dust levels, or geographical information can enhance classification accuracy. These models could be deployed to provide continuous assessment of the cleanliness of solar panels in real-time monitoring systems. Further optimization of the ensemble strategies, for example, by dynamic weighting of individual classifiers or meta-learning approaches, could be used to enhance the robustness of the framework. These may help improve further the performance, interpretability, and applicability of machine learning-based systems in the domain of solar energy management.